\documentclass[a4paper,12pt]{article}

\usepackage{amsmath}
\usepackage{amssymb}
\usepackage{dsfont}
\usepackage{bbm}
\usepackage{epsf}
\usepackage{epsfig}
\usepackage{graphicx}
\usepackage[english]{babel}
\usepackage[latin1]{inputenc}
\usepackage{rotating}
\usepackage{subfigure}
\usepackage{color}
\usepackage{slashed}
\usepackage{wrapfig}

\usepackage{txfonts}
\newcommand{\deltat}{a_{\text{t}}}
\newcommand{\trace}{\text{Tr} }

\newcommand{\epsf}{\epsilon^{1/4}}

\renewcommand{\vec}[1]{{\bf #1}}

\renewcommand{\epsilon}{\varepsilon}

\newcommand{\latteps}{\hat{\varepsilon}}

\begin{document}

\begin{center}
\begin{Large}
\textbf{Simulating plasma instabilities in SU(3) gauge theory} \\
\end{Large}

\medskip

J\"urgen Berges, Daniil Gelfand, Sebastian Scheffler, D\'enes Sexty \\

\textit{Institute for Nuclear Physics, Darmstadt University of Technology, 
Schlossgartenstr. 9, 64285 Darmstadt, Germany}\\
\end{center}

\begin{abstract}
We compute nonequilibrium dynamics of plasma instabilities in
classical-statistical lattice gauge theory in 3+1 dimensions. The 
simulations are done
for the first time for the SU($3$) gauge group relevant for quantum
chromodynamics. We find a qualitatively similar behavior as compared to
earlier investigations in SU($2$) gauge theory. The characteristic
growth rates are about $25$\% lower for given energy density, such that 
the isotropization process is slower. Measured in units of the 
characteristic screening mass, the primary growth rate is independent 
of the number of colors.
\end{abstract}

\section{Introduction}\label{Introduction}

The theoretical understanding of the apparent fast thermalization in
collision experiments of heavy nuclei at the Relativistic Heavy Ion
Collider provides a challenge for theory. It was noticed that plasma
instabilities in the locally anisotropic medium might play an
important role for a rapid isotropization of the equation of state,
which is relevant to explain the observed hydrodynamic
behavior~\cite{Plasmainst,WongYangMills,Romatschke:2006nk,Berges:2007re,oldsu3}.
Extensive studies have been carried out for SU($2$) pure gauge theory
using the hard-loop effective theory of soft excitations, which is
based on collisionless kinetic theory for hard particles coupled to a
soft classical field.\footnote{For numerical simulations which take
into account the backreaction of the soft fields on the hard particles
using a Boltzmann-Vlasov treatment see \cite{WongYangMills}. Studies
using transport or kinetic equations were also carried out
\cite{Xu:2007aa} in the spirit of the earlier bottom-up scenario
\cite{bottom-up-thermalization}, where no instability is present.}
This approach neglects quantum corrections and may also be considered
as an approximation of the classical-statistical field theory limit of
the respective quantum gauge
theory~\cite{Romatschke:2006nk,Berges:2007re}. Classical-statistical
lattice gauge theory provides a quantitative description in the
presence of sufficiently large energy density or occupation numbers
per mode. The simulations are done by numerical integration of the
classical lattice equations of motion and Monte Carlo sampling of
initial conditions.

In this work we present for the first time classical-statistical
lattice gauge theory simulations for the SU($3$) gauge group in 3+1
dimensions, relevant
for quantum chromodynamics (QCD).  Simulations using the hard-loop
approximation for SU(3) gauge group in 1+1 dimensions have been done
in Ref. \cite{oldsu3}.  Though the dynamics in one and three 
spatial dimensions are known to be very different, it is not expected 
that SU($3$) results will 
qualitatively change as compared to those previously obtained for SU(2), 
however
a quantitative estimate of the involved time scales seems imperative.
This work is a follow-up to our previous studies concerning the
numerically less demanding gauge group SU($2$)
\cite{Berges:2007re}. In this Letter we only describe the relevant
changes and refer to that reference for further computational
details. We find a qualitatively similar behavior as compared to the earlier
investigations in SU($2$) gauge theory, but for given initial energy
density the characteristic growth rates are about $25$\% lower such
that the isotropization process is slower.

The paper is organized as follows. In Section 2, we outline the setup
of our calculations, and the algorithm for solving the equations of motions
for the SU($3$) group. In Section 3 we present the results of the 
calculations and explain the dependence of growth rates on the 
gauge group in terms of diagrammatics. We conclude in
Section 4.

\section{Classical-statistical gauge field theory on a lattice}
\label{sec:classical-statistical}

Following Ref.~\cite{Berges:2007re} we use the Wilsonian lattice action on a 3+1 dimensional Minkowskian lattice,
\begin{eqnarray}\label{LatticeAction}
S[U] &=& - \beta_0 \sum_{x} \sum_i \left\{ \frac{1}{2 {\rm Tr}
\mathds{1}} \left( {\rm Tr}\, U_{x,0i} + {\rm Tr}\, U_{x,0i}^{\dagger}
\right) - 1 \right\}
\nonumber\\
&& + \beta_s \sum_{x} \sum_{\genfrac{}{}{0pt}{1}{i,j}{i<j}} \left\{ \frac{1}{2 {\rm
Tr} \mathds{1}} \left( {\rm Tr}\, U_{x,ij} + {\rm Tr}\, U_{x,ij}^{\dagger}
\right) - 1 \right\} \, ,
\end{eqnarray}
written in terms of plaquette variables
\begin{equation}
U_{x,\mu\nu} \equiv U_{x,\mu} U_{x+\hat\mu,\nu}
U^{\dagger}_{x+\hat\nu,\mu} U^{\dagger}_{x,\nu} \label{eq:plaq}\, ,
\end{equation}
where $U_{x,\nu\mu}^{\dagger}=U_{x,\mu\nu}\,$. Here $U_{x,\mu}$ denotes 
the link variable which is the
parallel transporter associated with the link from the neighboring
lattice point $x+\hat{\mu}$ to the point $x\equiv (t, \vec{x}) $ in 
the direction of
the lattice axis $\mu = 0,1,2,3$. The lattice parameters are defined as
\begin{equation}
\beta_0 \equiv \frac{2 \gamma {\rm Tr} \mathds{1}}{g_0^2} \,\, ,
\quad \beta_s \equiv \frac{2 {\rm Tr} \mathds{1}}{g_s^2 \gamma} \, ,
\label{eq:ganisoM}
\end{equation}
where $\gamma \equiv a_s/a_t$ is the ratio of 
the spatial and temporal lattice spacings, and we will consider $g_0 = g_s = g$ 
as the coupling constant of the lattice theory. 

Varying the action \eqref{LatticeAction} w. r. t.\ the spatial link variables 
$U_{x, j}$ yields the leapfrog-type equations of motion 
\begin{equation}\label{EOM}
\begin{split}
E^b_j(t , \vec{x}) \; = \;  E^b_j(t-\deltat, \vec{x}) - \,
\, \frac{2}{\gamma^2 a_s a_t g} \, \sum_{k} \,  \Bigl\{
&\textrm{Im} \trace \, \Bigl( \lambda^b U_{x, j} U_{(x +\hat{j}), k}
U^{\dagger}_{( x + \hat{k}), j} U^{\dagger}_{x, k} \Bigr) \\ 
 + \; & \textrm{Im} \trace \, \Bigl( \lambda^b U_{x,j} U^{\dagger}_{(x +
\hat{j} - \hat{k}) , k} U^{\dagger}_{(x - \hat{k}), j} U_{( x -
\hat{k}) ,k} \Bigr) \Bigr\} \, ,
\end{split}
\end{equation}
where $\lambda^b$ are the generators of SU($N$) satisfying $ \trace(
\lambda^a \lambda^b ) = \frac{1}{2} \delta^{ab} $. In the case of
SU($2$) these can be related to the Pauli matrices $\sigma^a$ through
$\lambda^a = \frac{1}{2} \sigma^a$. For SU($3$), the generators are
given by the Gell-Mann matrices divided by two.  The electric fields
in Eq.~\eqref{EOM} are
\begin{equation}\label{DefElectricField}
E^b_j(x) = \, \frac{2}{a_s \deltat g} \, \textrm {Im} \trace
\, \bigl( U_{x, 0j} \lambda^b \bigr) \, .
\end{equation}
Varying the action \eqref{LatticeAction} w. r. t.\ to a temporal link gives the Gauss constraint 
\begin{equation}\label{GaussLaw}
\sum_{j=1}^3 \, \bigl[ E_j^b(x) - U^{\dagger}_{x -\hat{j},j} \, E_j^b(x - \hat{j} \, ) U_{x -\hat{j},j} \, \bigr] =  0 \, . 
 \end{equation}

Using the gauge freedom the dynamics is computed in temporal axial
gauge, which is defined by the gauge condition $U_{x,0}
=\mathds{1}$. This choice still leaves the possibility of a
time-independent gauge fixing which we do not specify.
The algorithm for the numerical simulation may be summarized as follows: 
\footnote{Another simulation algorithm would be the 
Hamiltonian algorithm \cite{Kogut:1974ag}, which keeps track of the 
electric fields $E_j^b(x)$ in the Lie algebra space
instead of the link variables in group space. This 
way one would need to calculate exponentials of matrices to calculate
the time evolution of the spatial links.}

\begin{enumerate}
\item Using $E_j^b(t-\deltat,\vec{x})$ and $U_{(t,\vec{x}),j}$ the electric field is evolved to 
$E_j^b(t,\vec{x})$ with (\ref{EOM}).
\item The temporal plaquette $U_{(t,\vec{x}), 0j} \in \textrm{SU}(N)$ is calculated which satisfies \\
$ \textrm{Im} \trace ( U_{(t,\vec{x}), 0j} \lambda^b ) = {1\over 2} a_s a_t g E_j^b (t,\vec{x}) $.
\item The link variable $U_{(t+a_t,\vec{x}),j}$ is determined at the next time step 
from the defining equation of the temporal plaquette in temporal 
axial gauge: $ U_{(t+a_t,\vec{x}),j}=U_{(t,\vec{x}),0j}U_{(t,\vec{x}),j}$. 
\end{enumerate}
The main difference between the algorithms for $\text{SU}(2)$ and
$\text{SU}(3)$ occurs in step 2.  For $U \in \text{SU}(2) $ and given
$ \textrm{Im} \trace (U\sigma^a)= 2 b_a$, with $a = 1, 2, 3 $, one can
represent $U$ as 
$ U=\sqrt{1-b_a b_a} \mathds{1} + i b_a \sigma^a $ 
for the
solution which is close to the unit matrix.

For $\text{SU}(3)$ (and $\text{SU}(N)$ with $ N \ge 3$ in general) there is no simple corresponding procedure in step 2 and we solve for the temporal plaquette numerically.
For a general complex $3 \times 3$ matrix, which we use in the program, 
there are $18$ real variables ($3 \times 3$ complex numbers) and $28$ real, but 
not independent, equations 
(8 from (\ref{DefElectricField}), 2 from $ \textrm{det} U_{x,0j}=1$,
and 18 from $U_{x,0j}  U^{\dagger}_{x,0j} = \mathds{1}$). Since one only needs 18 equations, we use part of the unitarity equation:
components 12, 13, and 23, and the real part of components 11 and 22 
because the imaginary parts do not give any constraints on the variables. 
These equations are solved 
with the multi-dimensional Newton method \cite{numrec}. The fastest
simulation speed was achieved by using the unit matrix as a starting 
point of the Newton method, which typically leads to convergence in about 
$3$-$4$ steps. Since for every new link variable one has to iterate the Newton method
in 18 dimensions a few times, the $\text{SU}(3)$ simulation is considerably slower
than the corresponding $\text{SU}(2)$ simulation.

The initial Gaussian probability functional is chosen as in Ref.~\cite{Berges:2007re} such that
\begin{equation}\label{InitCond-1}
\langle \, \vert \, A^b_j(t=0,\vec{p}) \, \vert^2 \, \rangle \, = \, \frac{{\tilde A}^2}{(2 \pi)^{3/2} \Delta^2 \Delta_z} \, \exp \Bigl\{ -\frac{ p_x^2 + p_y^2 }{2 \Delta^2} - \frac{p_z^2}{2 \Delta_z^2} \Bigr\} \, , 
\end{equation}
where the gauge fields are 
calculated from the link variables using
\begin{equation}
A^b_j(t,\vec{x})= \frac{2}{ a_j g} \text{Im} \text{Tr} (U_{(t,\vec{x}),j} \lambda^b).
\end{equation}
We consider $E^a_j = - \dot{A_j^a} = 0 $ at $t=0$ fulfilling the
Gauss constraint~\eqref{GaussLaw}. 
We typically choose $ \Delta_z \ll \Delta$, and the distribution is practically  $\delta(p_z)$-like on the lattice.
Here $\Delta$ determines the typical
transverse momentum of the gluons and may be associated with the
saturation scale $Q_s$ at time $Q_s^{-1}$ in the saturation scenario~\cite{bottom-up-thermalization}. We solve the equations of motion with the aforementioned algorithm for a set
of initial configurations sampled according to Eq.~\eqref{InitCond-1}
and compute expectation values as averages over the results from the individual runs.

The local energy density in lattice units is determined by the action as
\begin{equation}\label{DefEnergyDensity}
\latteps (t, \vec{x}) \, \equiv \,  \frac{6 }{ g^2} \, \left( \gamma^2 \, \sum_j \,  1 -  \frac{ \trace \, U_{x,0j}
 + \trace \, U_{x,j0}}{6}
 + \sum_{j < k} \,  1 - \frac{ \trace \, U_{x, jk} 
  + \trace \, U_{x, kj}}{6}  \, \right)  \, 
\end{equation}
and we denote its average value by $ \latteps \equiv \langle \,
\latteps (t, \vec{x}) \, \rangle $. We choose the factor $\tilde{A}$
appearing in Eq.~\eqref{InitCond-1} such that $ \latteps = 0.05$. For
the conversion to physical units we fix the lattice spacing $a_s$ from
the relation between the physical average energy density and its
lattice analogue according to $\epsilon = \latteps \cdot a_s^{-4}$.
If $g$ is taken to be different from one then the lattice spacing is
altered by a factor $1 / \sqrt{g}$ , which follows from
Eq.~\eqref{DefEnergyDensity}.  The values for $a_s$ will not be
altered significantly as long as $g \sim \mathcal{O}(1)$. For later
reference, we note that the relation between $\latteps, N $ and
$\tilde{A}$ is approximately given by 
\begin{equation}\label{eq-en-dens}
 \latteps(t, \vec{x})  \,\approx \, \frac{a_s^4}{2} \, \sum_j \, \sum_{a=1}^{N^2 -1} \, \left\{ (E_j^a(t, \vec{x}))^2 + (B_j^a(t, \vec{x}))^2 \right\} 
\, \sim \, ( N^2 -1 ) \, \tilde{A}^2 \; 
\end{equation} for our initial conditions, 
where the magnetic fields are calculated from the spatial plaquettes according to 
\begin{equation}\label{DefMagneticField}
B^b_j(x) =  \epsilon^{jkl} \frac{1}{ a_s^2  g} \, 
\textrm {Im} \trace\, \bigl( U_{x, kl} \lambda^b \bigr) \, ,
\end{equation}
similarly to the electric fields (\ref{DefElectricField}).

\section{Results}
\label{sec:results}

\begin{figure}[t!]
\begin{center}
\epsfig{file=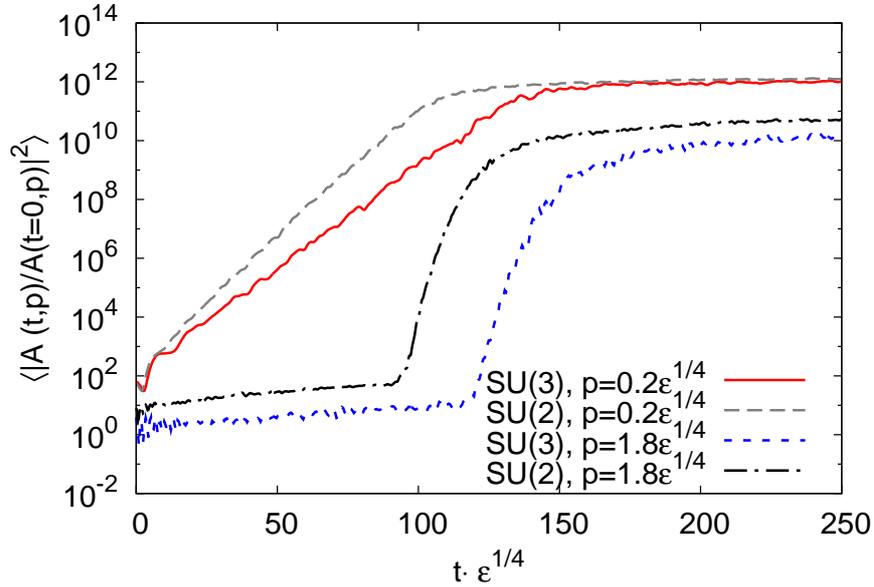, width = 12cm }
\caption{Fourier coefficients of the squared modulus of the gauge
field versus time for three different momenta parallel to the z-axis.
Compared is the time evolution of $\text{SU}(2)$ and $\text{SU}(3)$ gauge
fields for the same energy density $\epsilon$.}
\label{A-vs-t}
\end{center}
\end{figure}
Fig.~\ref{A-vs-t} shows the nonequilibrium time evolution of the
color-averaged squared modulus of two different Fourier coefficients
$A(t, \vec{p})$ of the gauge field in three spatial dimensions. They
are displayed as a function of time, normalized by the corresponding
field values at initial time, where time is measured in appropriate
units of the initial energy density $\epsilon$. The lattice size is
$64^3$ and the initial transverse width $\Delta =1.06 \,
\epsilon^{1/4}$. Here $\langle \, | A(t, \vec{p}) |^2 \, \rangle$ may be
associated to a particle number divided by frequency.

For comparison we show the SU($3$) results together with the
corresponding SU($2$) results. The behavior of both gauge groups is
qualitatively very similar. The plotted low-momentum modes clearly
show exponential growth starting at the very beginning of the
simulation. 
In contrast to these "primary" instabilities occurring at low momenta,
one observes from Fig.~\ref{A-vs-t} that gauge field modes at
sufficiently high momenta do not grow initially. The higher wave
number modes typically exhibit exponential growth at a "secondary"
stage that sets in later, but with a significantly larger growth rate.
The secondaries arise from fluctuation effects induced by the growth
in the lower momentum modes, which can be explained by taking into account
(2PI) resummed loop diagrams beyond the hard-loop approximation as
discussed in detail for SU($2$) gauge theory in
Ref.~\cite{Berges:2007re}. In that case it has been shown that the
exponential growth saturates when all loop diagrams become of order
one, which leads to a subsequent slow evolution towards a power-law
regime.\footnote{For related discussions in the context of scalar
inflaton dynamics in the early universe see
Refs.~\cite{Berges:2002cz,Berges:2008pc} and~\cite{Berges:2008mr} for
gauge fields.} Since the fastest growing mode in SU($2$) gauge theory
grows earlier to sizes where loop effects for higher modes become
important, one observes from Fig.~\ref{A-vs-t} that the secondary
growths also start earlier compared to the SU($3$) case.

\begin{figure}[t!]
\begin{center}
\epsfig{file=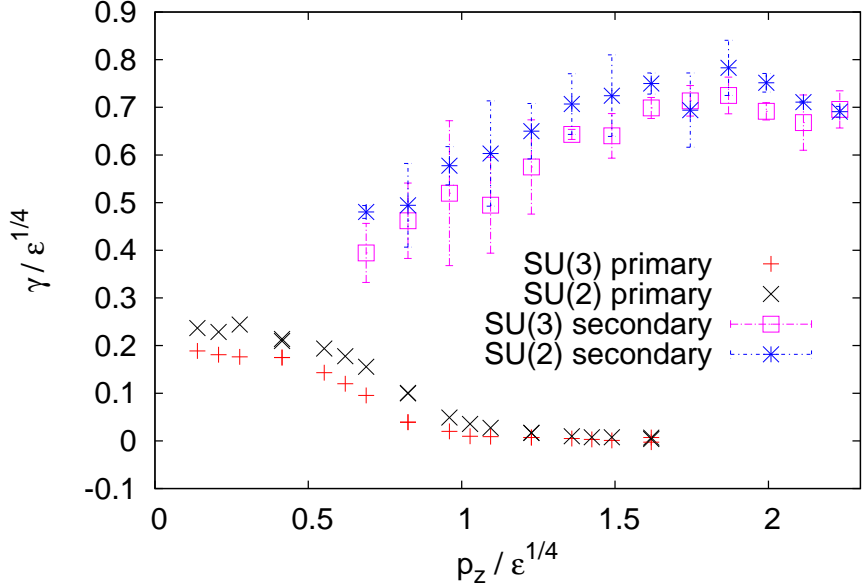, width = 12cm }
\caption{The primary and secondary growth rates for the
$\text{SU}(2)$ and $\text{SU}(3)$ gauge theory using the same energy density. The employed lattice sizes are $64^3$ and $96^3$. The exponential growth was fitted from an average of ca.\ 20 and ca.\ 100 runs for the 
$\text{SU}(3)$ and $\text{SU}(2)$ gauge groups, respectively.}
\label{rates}
\end{center}
\end{figure}
Fig.~\ref{rates} displays the momentum dependence of the growth rates for $\langle \, | A(t, \vec{p}) |^2 \, \rangle$ obtained from a fit to an exponential, which is done separately for the primary and secondary growth rates. One observes that while the primary
rates are approximately 25 \% bigger for $\text{SU}(2)$, the secondary
rates can be similar within the given errors.  
The dependence of the primary growth rates on the
gauge group can be understood from a diagrammatic analysis as follows. 
As outlined in Section~\ref{sec:classical-statistical}, the overall gauge field amplitude $\tilde{A}$ defined in (\ref{InitCond-1}) is chosen such that the average lattice energy density $\latteps$ has a prescribed value. For simulations of different gauge groups SU($N_1$) and SU($N_2$) at the same $\latteps$ using initial conditions sampled from~\eqref{InitCond-1} with identical $\Delta$ and $\Delta_z$, Eq.~\eqref{eq-en-dens} implies that the respective overall gauge field amplitudes $\tilde{A}$ are related according to
\begin{equation}\label{eq-a-tilde-ratio}
 (N_1^2 - 1 ) \, \tilde{A}_{\text{SU($N$}_1\text{)}}^2 \,=\,  (N_2^2 - 1 ) \, \tilde{A}_{\text{SU($N$}_2\text{)}}^2 \; .
\end{equation}
Plugging in the relevant numbers we find for SU(2) and SU(3)
\begin{equation}\label{eq-ratio-Atilde-SU2-SU3}
\tilde{A}^2_{\text{SU($3$)}} = \frac{3}{8} \, \tilde{A}^2_{\text{SU($2$)}} \; .
\end{equation}

\begin{figure}[t!]
\begin{center}
\epsfig{file=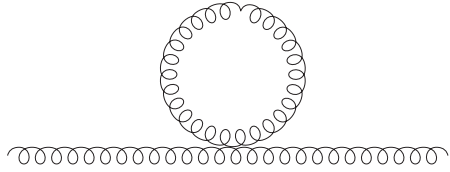, width = 6cm }
\caption{The tadpole diagram whose non-vacuum part determines $m_T$ as described in the main text.}
\label{tadpole}
\end{center}
\end{figure}
\begin{figure}[t!]
\begin{center}
\epsfig{file=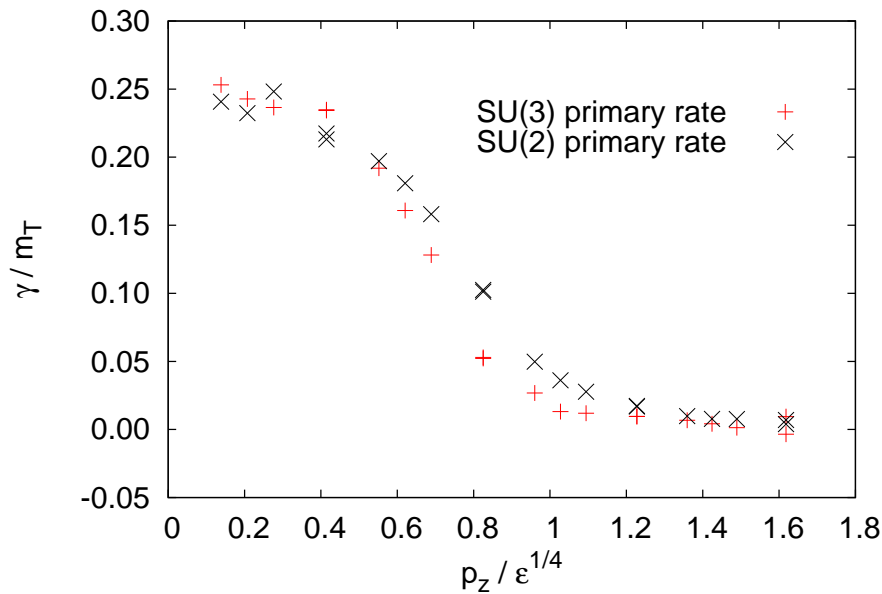, width = 12cm }
\caption{Comparison of classical-statistical simulation results for the primary growth rates
as a function of momentum for SU($2$) and SU($3$) gauge theories, measured in units of the respective $m_T$. The same parameters are used as in Fig.~\ref{A-vs-t}.}
\label{fig:rescaled}
\end{center}
\end{figure}

In Ref.~\cite{Berges:2007re} we have analyzed for SU($2$) gauge theory relevant diagrammatic contributions to the nonequilibrium evolution at various characteristic time scales, which holds along the same lines for SU($3$). As a characteristic self-energy contribution relevant for the primary growth rates at early times, we consider the non-vacuum part $m^2_T$ of the tadpole diagram shown in Fig.~\ref{tadpole}. For SU($N$) gauge theory the parametric $N$-dependence of that contribution is
\begin{equation}\label{eq-minf-N-dependence-1}
m^2_{T,\text{SU($N$)}}  \, \sim \, N \, \tilde{A}^2_{\text{SU($N$)}} \; , 
\end{equation}
which implies
\begin{equation}\label{eq-minf-N-dependence-2}
 \frac{m^2_{T,\text{SU($N_1$)}}}{m^2_{T,\text{SU($N_2$)}}} \, = \, \frac{N_1}{N_2} \, \frac{\tilde{A}^2_{\text{SU($N_1$)}}}{\tilde{A}^2_{\text{SU($N_2$)}}}
\end{equation}
at fixed $\latteps$. Using Eq.~\eqref{eq-ratio-Atilde-SU2-SU3} we can thus compare the change in $m_T$ when going from SU($2$) to SU($3$) at fixed energy density which amounts to
\begin{equation}\label{eq-m-change}
 m_{T,\text{SU($3$)}} \,=\, \frac{3}{4} \, m_{T,\text{SU($2$)}} \; .
\end{equation}
Fig.~\ref{fig:rescaled} shows that the classical-statistical
simulations for SU($2$) and SU($3$) gauge theory give indeed
approximately the same results for primary growth rates in units of
the respective $m_T$.\footnote{This mass may be directly related to
$m_\infty$ in the 'hard loop' approximation, calculated from the
anisotropic distribution of the hard modes~\cite{Plasmainst}. Note
that for our case $m_T$ does not scale with $N$ like the thermal mass,
because we keep the energy density fixed, which results in different
temperatures for theories with different numbers of degrees of freedom.
}

In order to obtain an estimate in physical units one may consider an initial energy density of about $5$--$25$ GeV/fm$^3$ for RHIC experiments, and a projected factor of about two more for LHC energies. The inverse of the maximum primary growth rate for $|\, A(t, \vec{p}) \, |^2$ then leads to the characteristic time scales 
\begin{eqnarray}\label{inverse-primary-gr}
 \gamma_{\text{max.\ pr.}}^{-1} & \simeq & 1.6\ -\ 2.4 \, \text{fm}/\text{c} \hspace{1cm}   (\text{RHIC}) \, , \\ 
 \gamma_{\text{max.\ pr.}}^{-1} & \simeq & 1.3\ -\ 2.0 \, \text{fm}/\text{c} \hspace{1cm}   (\text{LHC})\, .
\end{eqnarray}
The results are rather insensitive to the precise value of the initial
energy density because they scale with the fourth root of
$\epsilon$. For comparison, a time scale associated with the largest
observed secondary growth rate is about a factor of three shorter than
what is given in (\ref{inverse-primary-gr}). However, even though
secondaries can reach considerably higher growth rates than primaries,
they start later. As a consequence, a certain range of higher momentum
modes can 'catch up' with initially faster growing infrared modes
before the exponential growth stops, as seen in
Fig.~\ref{A-vs-t}. This leads to a relatively fast effective
isotropization of a finite momentum range, while higher momentum modes
do not isotropize on a time scale characterized by plasma
instabilities.
 
\begin{figure}[t!]
\begin{center}
\epsfig{file=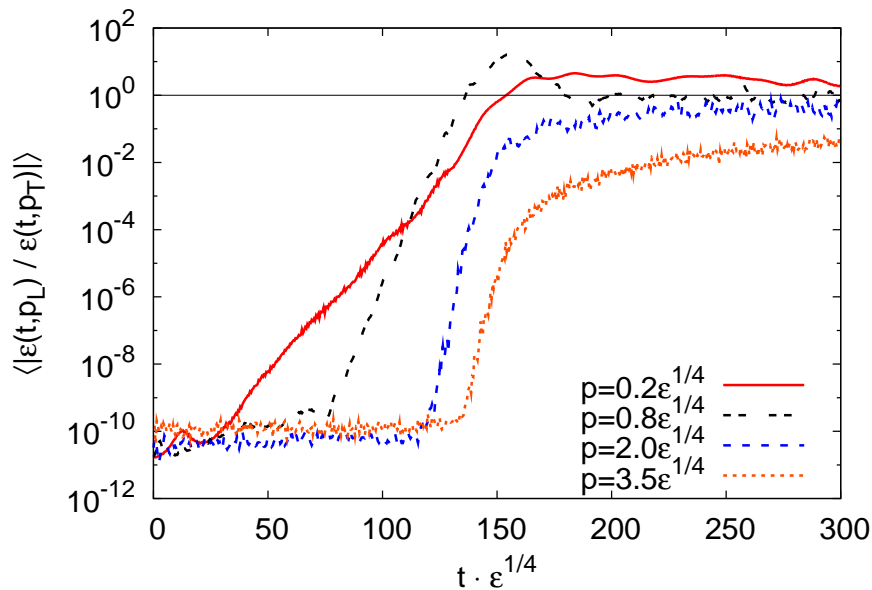, width = 12cm }
\caption{ 
The ratio of the longitudinal and transverse modes of the Fourier transform 
of the spatially dependent energy density for various momenta, measured on 
$N^3=64^3$ lattices, using $\text{SU}(3)$ gauge group.
 }
\label{fig:pressure}
\end{center}
\end{figure}

We study the process of isotropization in terms of the local energy
density~\eqref{DefEnergyDensity} which is a gauge invariant
quantity. For this, we Fourier transform $\latteps(t, \vec{x})$
with  respect to $\vec{x}$ and compute the absolute value of the ratio
\begin{equation}\label{eq-edens-ratios}
\frac{\latteps(t, \vec{p}_L ) }{\latteps(t, \vec{p}_T)} \; \textrm{with } \, \vec{p}_L \parallel \hat{z} \, , \, \vec{p}_T \perp \hat{z} \, , \, \, \textrm{and} \, \, | \ \vec{p}_L \, | = |\, \vec{p}_T \, | \; .
\end{equation}
If at some time $t$ the system reaches an isotropic state the mean absolute value of~\eqref{eq-edens-ratios} has to be one for all momenta.  The time evolution of~\eqref{eq-edens-ratios} is depicted in Fig.~\ref{fig:pressure} for several momenta. Similarly to the $\text{SU}(2)$ case one observes that the low-momentum modes of the energy density isotropize, while the high-momentum modes are still anisotropic after the instability growth ends. 

Inspecting Fig.~\ref{fig:pressure} one sees that the Fourier coefficients of $\latteps$ become isotropic up to approximately $| \, \vec{p} \, | \simeq 1 - 2 \, \epsf $ at the time of saturation. Using the same values for the physical energy density $\epsilon$ as for~\eqref{inverse-primary-gr} one finds effective isotropization up to a characteristic momentum of about
\begin{eqnarray}
| \, {\bf p} \, | & \lesssim & 1 \, {\rm GeV} \, ,
\end{eqnarray}
which approximately agrees with the case of the SU(2) gauge theory.

\section{Conclusions}

Nonabelian plasma instabilities in SU($3$) gauge theory relevant for QCD exhibit a qualitatively similar behavior as previously observed for the SU($2$) group. The main quantitative differences concern the reduction of primary growth rates by about 25 \% for given initial energy density. These differences can be related to the parametric dependence on the number of colors of one-loop corrections to the self-energy. As a consequence, we find that the properly rescaled primary growth rates agree even quantitatively rather well for the different gauge groups. Though the nonlinear dynamics underlying the secondary growth rates leads to remaining quantitative differences between SU($3$) and SU($2$), this fact is less important for phenomenology. The primary growth rates determine the characteristic time scales for isotropization at low momenta, which turns out to be (even) slower than previously suggested by SU($2$) results.    

\vspace*{0.5cm}

\noindent
This work is supported in part by the BMBF grant 06DA267, and by the
DFG under contract SFB634. Part of this work was inspired by the program 
on "Nonequilibrium Dynamics in Particle Physics and
Cosmology" (2008) at the Kavli Institute for Theoretical Physics in Santa Barbara,
supported by the NSF under grant PHY05-51164.

\end{document}